\documentstyle[graphicx]{mn}

\newcommand{\etal}{et al.\ }

\newcommand{\eg}{e.g.\ }
\newcommand{\GC}{Galactic Centre}
\newcommand{\GCR}{Galactic Centre Region}

\newcommand{\NH}{$N_{\rm H}$}
\newcommand{\NHUNIT}{H~cm$^{-2}$}
\newcommand{\FLUXUNIT}{erg~s$^{-1}$~cm$^{-2}$}
\newcommand{\LUMIUNIT}{erg~s$^{-1}$}

\newcommand{\XMM}{\it XMM}
\newcommand{\XMMN}{\it XMM-Newton}
\newcommand{\Chandra}{\it Chandra}

\newcommand{\HII}{\protect \hbox{H\,{\sc ii}}}

\title{The discovery of a new non-thermal X-ray filament near the Galactic 
Centre}
\author[M. Sakano, R.S Warwick, A. Decourchelle and P. Predehl]
       {M. Sakano,$^1$\thanks{Japan Society for the Promotion of Science
       (JSPS).} R. S. Warwick$^1$, A. Decourchelle$^2$ and P. Predehl$^3$\\
        $^1$Department of Physics and Astronomy, University of Leicester,
	Leicester LE1 7RH, UK\\
	$^2$CEA/DSM/DAPNIA, Service d'Astrophysique, C.E. Saclay, 
	91191 Gif-sur-Yvette Cedex, France\\
	$^3$Max-Planck-Institut f\"ur extraterrestrische Physik,
	Postfach 1312, D-85741 Garching, Germany}

\date{Accepted XXXX. Received XXXX 2002}

\pagerange{\pageref{firstpage}--\pageref{lastpage}}
\pubyear{2002}

\begin{document}

\maketitle

\label{firstpage}

\begin{abstract}
We report the discovery by {\XMMN} and {\Chandra} of a hard extended 
X-ray source (XMM~J174540$-$2904.5) associated with a compact non-thermal 
radio filament (the Sgr A-E `wisp'=1LC 359.888$-$0.086=G359.88$-$0.07), which is
located within $\sim 4$ arcmin of the {\GC}. The source position is 
also coincident with the peak of the molecular cloud, M~$-$0.13$-$0.08 
(the `20~km~s$^{-1}$' cloud).  The X-ray spectrum is non-thermal with 
an energy index of 1.0$^{+1.1}_{-0.9}$ and column density of 
{$38^{+7}_{-11}\times 10^{22}${\NHUNIT}.  The observed 2--10 keV flux 
of 4$\times 10^{-13}${\FLUXUNIT} converts to an unabsorbed X-ray
luminosity of 1$\times 10^{34}${\LUMIUNIT} assuming a distance of 8.0~kpc.
The high column density strongly suggests that this source is located in
or behind the {\GCR}.  Taking account of the broad-band spectrum, as
well as the source morphology and the positional coincidence with a 
molecular cloud, we concluded that both the radio and X-ray emission 
are the result of synchrotron radiation.  This is the first time
a filamentary structure in the {\GCR} has been shown, unequivocally,
to have a non-thermal X-ray spectrum.}

\end{abstract}

\begin{keywords}
Galaxy:~centre -- X-rays:ISM -- X-rays:individual:XMM~J174540$-$2904.5

\end{keywords}

\section{Introduction}

The inner $\sim 300$ pc of our galaxy is a unique region.  The compact radio 
source Sgr$\rm~A^{*}$, which is spatially coincident with the dynamical centre 
of the Galaxy, marks the presence at the Galactic nucleus of a 3$\times
10^{6}$ solar mass, black-hole (Genzel {\etal} 2000; Ghez {\etal} 2000;
Eckart {\etal} 2002).  The bolometric luminosity of this
super-massive object is extraordinarily low at the present time, although  
low-level X-ray flaring activity has 
been observed in X-rays on at least two recent occasions (Baganoff \etal
2001; Goldwurm \etal 2002). The {\GCR} as a whole embodies a dense mix of 
molecular clouds, star clusters, {\HII} regions, 
supernova remnants, hot plasma, energetic particles and magnetic fields
({\eg} see the review by Mezger, Duschl, \& Zylka 1996).  Arguably the most 
striking large-scale 
structures are the non-thermal filaments (hereafter NTF) seen in radio 
observations, some of which extend over scales of tens of parsecs, run 
largely perpendicular to the Galactic plane and, presumably, trace regions 
of either enhanced magnetic field and/or a local source of relativistic 
particles ({\eg} Morris 1994; Mezger {\etal} 1996; Lang, Morris, \& Echevarria 1999). 
At the other end of the electromagnetic spectrum, the most obvious feature 
of X-ray images 
(Sakano \etal 2002; Wang, Gotthelf \& Lang 2002; Warwick 2002), in addition to a
population of luminous 
($\rm  L_X \ga 10^{36} ~erg~s^{-1}$) X-ray binaries, is a wide-spread 
distribution of hot ($>1$~keV), largely thermal plasma.

Here we combine recent X-ray observations of the {\GCR} by {\XMMN} and 
{\Chandra} with published radio data to identify a hard X-ray source 
coincident with a  non-thermal radio filament. In general the very large 
scale NTFs seen in the radio band do not have obvious X-ray counterparts. 
However, there is evidence of X-ray emission in at least one compact NTF 
identified in radio observations (Wang {\etal} 2002; Wang 2002). Arguably, the present 
discovery provides the clearest example of an X-ray bright, compact NTF. 

\section[]{The X-ray Observations}

A major survey of the region along the Galactic Plane within $1^{\circ}$ of 
the {\GC} is in progress with {\XMMN}. The full programme consists of a set
of 10 overlapping pointings (plus one or two additional observations 
targeted at specific Galactic Centre sources). A preliminary, mosaiced
image of the whole survey region after correction for the instrument 
background and exposure variations, including telescope vignetting, has 
been presented in Warwick (2002).

\begin{figure*}
\centering
\begin{minipage}{0.67\textwidth}
\centering\hbox{\includegraphics[width=12 cm, angle=270]{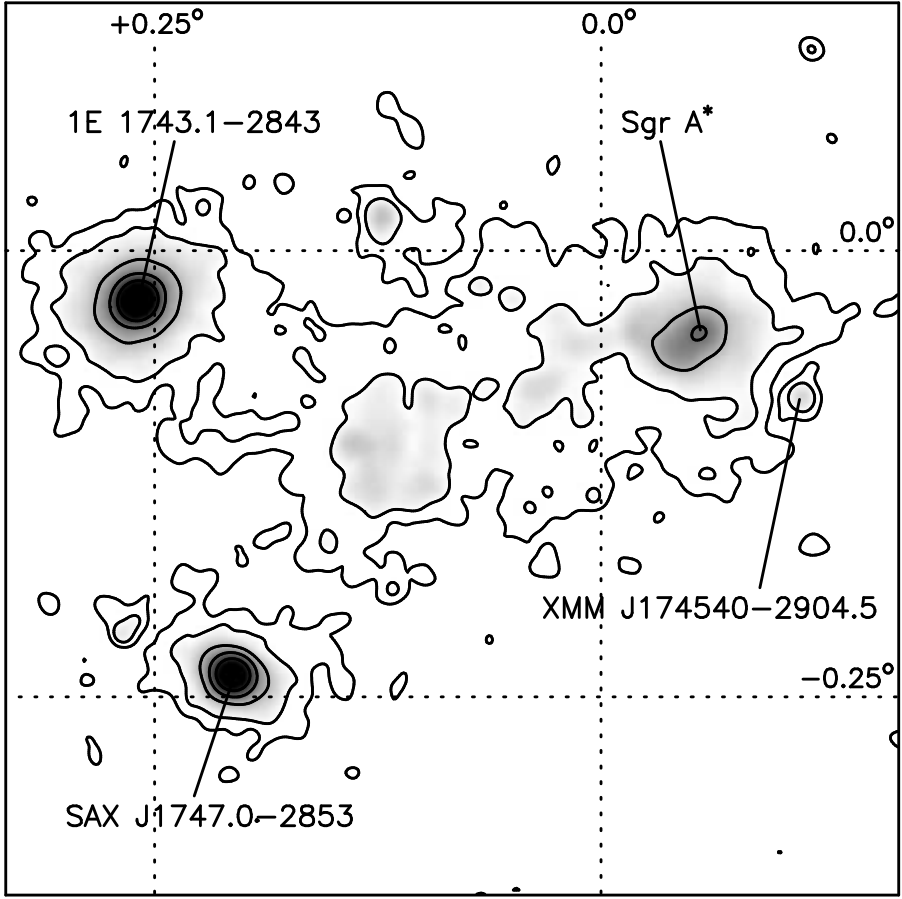}}
\caption{An extract from the {\XMMN} mosaiced image of the Galactic Centre 
Region based  on the combined MOS 1/2 datasets. The selected bandpass is 
2--9 keV.  The grid shows Galactic coordinates. The $0.5\times 0.5$~deg$^2$
field encompasses the luminous discrete sources 
1E 1743.1$-$2843 and SAX J1747.0$-$2853 together with the bright extended 
X-ray emission associated with the Sgr A East complex.  Diffuse X-ray 
emission also runs eastwards from Sgr~A East into the so-called Radio Arc 
region. The position of XMM~J174540$-$2904.5, 4~arcmin to the south-west
of Sgr$\rm~A^{*}$ is also indicated.
}
\label{xmm_image}
\end{minipage}~~
\end{figure*}

The current paper focuses on one of the observations from the {\XMMN} {\GC}
programme  targeted at the Sgr A complex. This observation (designated GC6)
was carried out by {\XMMN} on September 4, 2001. Here we concentrate on the 
measurements made 
by the {\XMMN} EPIC cameras, which consist of two MOS CCD units 
(Turner \etal 2001) and one pn CCD (Str\"{u}der \etal 2001).  For the {\GC} 
programme the EPIC MOS and pn cameras  were operated in {\it Full 
Frame Mode} and {\it Extended Full Frame Mode}, respectively, with the medium 
filter selected. The data reduction and filtering were carried out using the 
Standard Analysis Software ({\sc SAS}), Version 5.2. Although the instrument 
background was relatively stable during the GC6 observation, we rejected 
some time intervals based on the light curve of the full-field data above 
10~keV.  The resulting effective exposure times for MOS1, MOS2 and pn 
cameras were 22.4~ks, 24.0~ks, and 17.5~ks, respectively. For the MOS 
cameras we utilised pixel patterns of 0--12 (single to quadruple), whereas  
for the pn camera, we accepted only single (pattern 0) events\footnote{This
conservative approach was adopted  because of calibration uncertainties 
pertaining at the time of the analysis to other pn event types recorded  
in {\it Extended Full Frame Mode}.}.

We have also employed data from a {\Chandra} observation of this region 
made on July 8, 2000 with the Advanced CCD Imaging Spectrometer (ACIS-I) 
in {\it VFAINT Mode}.  In this case, data reduction and filtering 
were carried out with the Chandra Interactive Analysis Software
({\sc CIAO}), Version 2.2.  After applying standard filtering criteria, the 
effective exposure time was 38.7~ks.

\section[]{Results}

\subsection[]{X-ray images and the radio counterpart\label{sec:image}}

\begin{figure}
\centering
\centering\hbox{\includegraphics[width=8 cm]{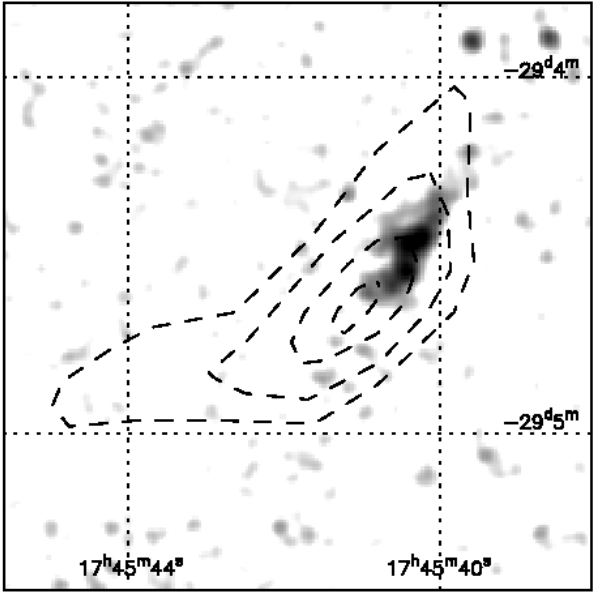}}
 \caption{The {\Chandra} image of XMM~J174540$-$2904.5 in the 2--8 keV
 band.  It is smoothed with a Gaussian filter with the sigma of 1
 arcsec.  The contours correspond to 2 cm radio continuum 
measurements by VLA with the angular resolution of 11\farcs8$\times$9\farcs8,
and are taken from Fig.~2 of Ho \etal (1985); here we reproduce 
only every other contour from the original radio map.}
  \label{cxo_image}
\end{figure}

Fig. \ref{xmm_image} shows an extract from the {\XMMN} mosaic of the 
{\GCR}. Our attention was drawn to a discrete source seen in this 
image 4~arcmin to the south west of Sgr$\rm~A^{*}$ by virtue of its extremely 
hard X-ray spectrum. The position of this source based on the 
{\XMMN} data is (17$^{\rm h}$~45$^{\rm
m}$~40\fs4, $-$29\degr~04\farcm5) in J2000 coordinates with an 
error radius of 6~arcsec, and was accordingly designated as 
XMM~J174540$-$2904.5.  The source was found to have an extent slightly 
greater than the point-spread function (PSF) of the {\XMMN} mirror, 
implying an angular size greater than $\sim 10$ arcsec. This was confirmed by 
reference to the {\Chandra} image (Fig.~\ref{cxo_image}), where 
the higher spatial resolution clearly delineates an elongated morphology
on an angular scale of $\sim$15 arcsec. In the case of {\Chandra} the 
astrometric accuracy of the X-ray data is better than $\sim 1$ arcsec.

The position of XMM~J174540$-$2904.5 coincides with that of an
extended non-thermal radio source detected in the 2 and 6 cm wavelengths.
Ho \etal (1985) refer to this
source as Sgr~A-E `wisp' and in Fig.~\ref{cxo_image} we 
have overlaid several of the radio contours from the 2 cm continuum 
radio map shown in Fig. 2 of the Ho \etal paper. Interestingly
the X-ray emission originates in a much more compact region
than the radio emission and the peak in the X-ray surface brightness
is offset from the corresponding feature in the radio image.
Nevertheless the association of the radio and X-ray sources looks
secure given the close alignment of the X-ray and radio
elongation in the region of overlap.

More recently, Lazio \& Cordes (1998) have catalogued the radio source as 
1LC 359.888$-$0.086 (17$^{\rm h}$~45$^{\rm m}$~41\fs131, 
$-$29\degr~04\arcmin~36\farcs43, J2000) and noted that at 1281 MHz it has
an angular extent of 47~arcsec, whereas the instrumental resolution is
5~arcsec.  Lang {\etal} (1999) confirmed this result with a 20~cm
observation and designated it as G359.88$-$0.07.
This position also
corresponds to the peak position of a giant molecular cloud
M~$-$0.13$-$0.08 (the `20~km~s$^{-1}$' cloud) (Mezger \etal 1986).

Combining the radio measurements at 15~GHz (2~cm), 5~GHz (6~cm) and 1.3~GHz 
(Ho {\etal} 1985; Lazio \& Cordes 1998), we find all the three data points
to be well fitted with a power-law model with the spectral index
$\alpha\approx 0.4$ ($I_{E} \propto E^{-\alpha}$).

\subsection[]{X-ray spectra \label{sec:spec}}

We extracted source spectra from the {\XMM}-EPIC MOS and pn 
data using a 16~arcsec radius source cell centred on 
XMM~J174540$-$2904.5. Corresponding background spectra
were taken from a concentric annulus  with an inner and outer
radius  of 32 and 48 arcsec respectively.  For the
{\Chandra} ACIS data, we chose an elliptical source region with major and
minor axis diameters of 36 and 15 arcsec and a concentric background 
region bounded by the source region and an outer dimension twice its size.
The background-subtracted spectra were analysed using {\sc xspec}
version 11.1.0.  Initially we tried fitting the {\XMM} MOS~1, 2, 
and pn data simultaneously and also, in a separate exercise,
the {\Chandra} data alone. However, the results from the two observatories
were quite consistent except for some minor normalisation differences. 
Therefore we next adopted the approach of simultaneously fitting all four 
datasets, but allowing the global normalisations for the {\XMM} and 
{\Chandra} data to be free.

We investigated both power-law continuum and thin-thermal plasma 
emission models, modified by absorption, and found
that the plasma model was not appropriate because the 90~per cent lower-limit
of the temperature was very high, {\it i.e.} $\sim 40$~keV.  
Note that we assume the solar abundance ratios (Anders \& Grevesse
1989).  Here we apply the {\sc wabs} model, which is based on the cross
section of Morrison \& McCammon (1983) and the above-mentioned solar abundances.

\begin{figure}
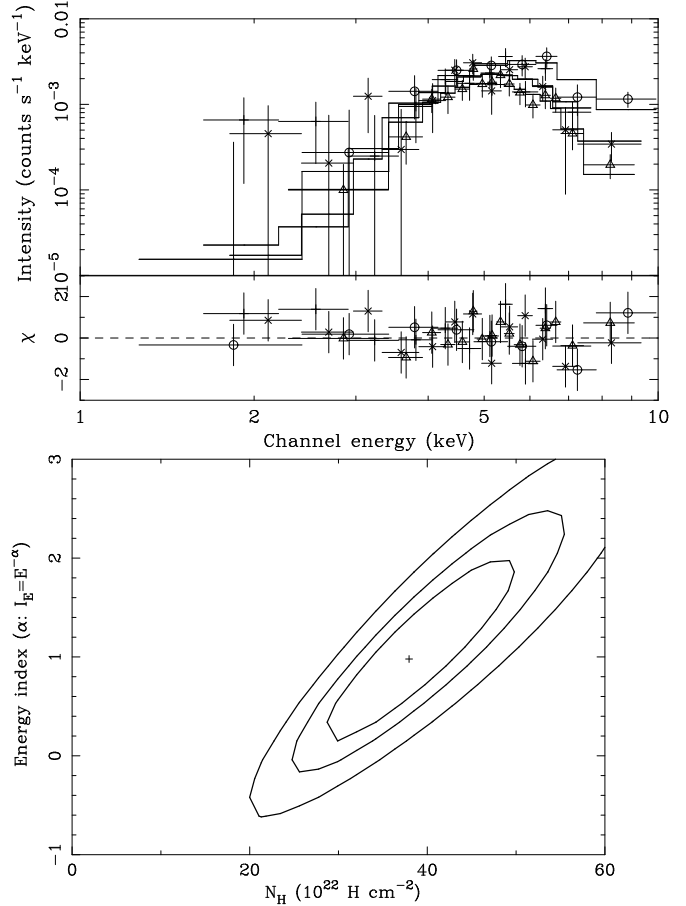

\hbox{\includegraphics[width=6 cm, angle=270]{allexhsrc_pow_gain_rev.ps}}
\hbox{\includegraphics[width=6 cm, angle=270]{allexhsrc_pow_gain_contalph.ps}}
 \caption{The X-ray spectral properties  of XMM J174540$-$2904.5. {\it Upper
panel}: The measured X-ray count rate spectra taken with {\XMM} and
{\Chandra}, together with the best fitting absorbed power-law model and
the fitting residuals. 
{\it Lower panel:} The confidence contours in $\alpha$ (energy index) versus {\NH} space --
the three contours represent respectively 68, 90 and 99\% confidence limits.  
}
 \label{fig:exhsrc}
\end{figure}

Fig.~\ref{fig:exhsrc} shows the observed spectra together with
the best-fitting power-law model and the corresponding
confidence contour for column density versus energy index. 
Table~\ref{tbl:fit} summarises the 
results from both the separate and combined {\XMM} 
and {\Chandra} analysis.  The best-fitting
parameters for the power-law model are an energy index of 1.0 (0.1--2.1),
a hydrogen column density of 38 (27--45) $\times 10^{22}${\NHUNIT},
and a 2--10 keV observed X-ray flux of 4$\times 10^{-13}${\FLUXUNIT}.
The latter converts to a 2--10 keV unabsorbed luminosity of 1$\times
10^{34}${\LUMIUNIT} assuming a source distance of  8.0~kpc. 
We found out that the inclusion of dust scattering does not affect significantly
the estimate of the absorption column or flux (This is based on the
inclusion in the spectral fitting of the {\sc dust} model in {\sc xspec} 
using parameters estimated from the work of Mitsuda {\etal} 1990).
We also tried a model including a narrow line at 6.4 keV from neutral iron.
The line is, however, not detected at a statistically significant level; 
the derived upper limit of the equivalent width is 280~eV.

Further checks revealed no significant flux or spectral variations during
the observation.

\begin{table*}
 \centering
 \begin{minipage}{140mm}
  \caption[]{The best-fitting spectral parameters. \label{tbl:fit}
}
  \begin{tabular}{ccccl}
   \hline
    & {\XMM} & {\Chandra} & {\XMM}+{\Chandra} & Unit\\
   \hline
   {\NH} & 33 (20--48) & 37 (24--64) & 38 (27--45) & ($10^{22}${\NHUNIT})\\
   $\alpha$\footnote{Energy index $\alpha$, where $I_E\propto E^{-\alpha}$.} & 0.7 ($-$0.7--2.2) & 0.7 ($-$0.3--3.3) & 1.0 (0.1--2.1) & \\
   $F_{\rm X}$ & 3.8  &  4.2  & 4.1/3.6\footnote{{\XMM}/{\Chandra}} & ($10^{-13}${\FLUXUNIT}; 2--10 keV)\\
   \hline
\end{tabular}
\medskip

Numbers in parentheses are 90\% confidence uncertainties for one
  interesting parameter.
\end{minipage}
\end{table*}

\section[]{Discussion}

The spatial coincidence and the similar orientations of the extended
morphology (see Section~\ref{sec:image}), provide strong evidence for the 
association of the X-ray and radio sources.  The heavy absorption in the 
X-ray band further implies that this source is embedded in or beyond the 
molecular cloud M$-$0.13$-$0.08, which Zylka, Mezger \& Wink (1990),
on the basis of infrared to radio observations, propose 
is actually located in front of the {\GC} (Sgr~A$^*$) by several tens 
of parsecs.  In addition, the relatively weak 6.4-keV
line from the source suggests that the surrounding cloud column density 
is less than 3$\times 10^{23}${\NHUNIT} (for an isotropic
cloud; see Inoue 1985), although, of course, the source need not be 
at the cloud centre and may in fact be behind it.  Coil \& Ho (2000)
argue from their radio observations that the radio `wisp' may be part of 
a SNR (designated as G~359.92$-$0.09) which lies to the side of and
behind the M$-$0.13$-$0.08 cloud.

The flat X-ray spectrum implies that the emission is non-thermal 
in origin. Non-thermal X-ray emission has previously been detected 
from a number of Galactic SNRs, ({\eg} Koyama {\etal} 1995), 
and is generally interpreted as a synchrotron emission of
relativistic electrons with energy of up to $\sim$100~TeV ({\eg}
Reynolds \& Keohane 1999).  Similarly the acceleration 
of electrons up to $\sim$100~TeV is also observed in 
pulsar wind nebulae, such as in the Crab nebula ({\eg} Aharonian
{\etal} 2000).

The lifetime of synchrotron-emitting electrons is inversely proportional 
to the square of the magnetic field.  The observed magnetic field in 
the {\GCR} is high, typically $\sim$1~mG ({\eg} Morris 1994), although it is
not known whether this strong magnetic field pervades the whole {\GCR}
or is more localized.  Specifically in the case of the
Sgr~A-E `wisp', Ho \etal (1985) estimated based on an
equipartition argument that the magnetic field is 0.3~mG. The lifetime
of the $\sim 20$ TeV electrons emitting hard X-rays through synchrotron 
radiation is
then $\sim$7~yr.  This lifetime is comparable to the observed spatial extent of
the filament which is $\sim$2~light year in the X-ray band. Since the source
was detected at a similar flux level on separate occasions by 
{\Chandra} and {\XMMN}, spaced by a time interval of 1~yr, we can conclude
that there is continuous injection (or re-acceleration) of high-energy 
electrons.

If we assume a constant injection of electrons with an energy
spectrum $\propto E^{-p}$, starting some time $t$ earlier, then the emitted 
synchrotron spectrum is $\propto tE^{-\frac{p-1}{2}}$ in the low
energy band ({\eg} radio) and $\propto E^{-\frac{p}{2}}$ in the high energy
band ({\eg} X-ray) (see Longair 1994).  In other words the spectral 
index in the X-ray band should be 0.5 steeper than that measured
in the radio band, with a spectral break lying somewhere between the 
two. The spectral indices measured in the radio ($\alpha\sim 0.4$) and
X-ray ($\alpha\sim 1.0$) bands are, in fact,
quite consistent with this model, although other 
interpretations are possible due to the relatively large 
measurement errors.  Nevertheless this is further support for
the synchrotron hypothesis.

The spatial coincidence of the filament with the peak of a molecular cloud 
adds support to the conjecture that the source is located in the
{\GCR}. This assumes that the presence of the non-thermal 
emission is in some way related to the high-density conditions within  
the molecular cloud, even if the mechanism for the acceleration of particles 
in such an environment is not clear. Presumably the X-ray emission locates 
the acceleration site whereas the more extended radio source
maps the extent to which $\sim 1$ GeV electrons diffuse
through a region of enhanced but (according to the
radio polarimetry observations of Ho {\etal} 1985) somewhat tangled 
magnetic field. 

Ho {\etal} (1985) suggest that the radio `wisp' (corresponding to XMM
J174540$-$2904.5) traces the edge of a possible SNR which is located to
the south of and presumably just in front of Sgr~A East.  More recently, Coil \& Ho
(2000) have found an intriguing association between the radio 20~cm continuum
and NH$_3$ line emission at the location of the `wisp' and argue that
the radio `wisp' may coincide with the edge of a SNR.  XMM
J174540$-$2904.5, accordingly, may be at a shock front in the compressed
high-density cloud and the site of the particle acceleration.

There are about 10 (isolated) radio filamentary structures known in the {\GCR}
({\eg} Morris 1994; LaRosa, Lazio \& Kassim 2001), most of which show 
extended morphologies aligned perpendicular to the Galactic Plane, as is 
the case for the present filament. On the other hand, the detection
of X-ray emission associated with such radio filaments is still rare. 
The best previous example of a radio and X-ray correlation is that
of NTF G359.54+0.18 (Yusef-Zadeh, Wardle \& Parastaran 1997) for which
Wang (2002) has reported the detection by {\Chandra} of an X-ray thread
about 1~arcmin long but only $\sim$1~arcsec wide.
Koyama (2001), Yusef-Zadeh {\etal} (2002), and Bamba et al. (2002) have also reported the 
detection of a number of other X-ray filaments in {\Chandra} data,  
although detailed studies have not been possible due to limited count 
statistics and because corresponding radio filamentary structures have
not been identified.  This is the first clear detection of the X-ray
filament which has a radio counterpart and unequivocally has a
non-thermal X-ray spectrum, although it might be a part of a large SNR.

Finally we note that although our preferred interpretation of 
XMM~J174540$-$2904.5 is that it is the X-ray counterpart of a 
non-thermal filamentary structure in the {\GCR}, there are other 
possible scenarios for its origin. For example, it could be 
an extragalactic background object, namely a one-sided jet 
emanating from a QSO, although the lack of a central point source 
corresponding to the QSO core somewhat weakens this argument.
In this context detailed investigation of the X-ray morphology 
via deep {\Chandra} imaging and high-resolution mapping in the 
radio band are the way forward.

\section*{Acknowledgments}

The authors would like to express their thanks to all those who have 
contributed to the successful development and operation of {\XMMN}. In 
addition we should like to acknowledge the help of many colleagues at 
Leicester, especially R. Griffiths, R. Saxton, S. Sembay, I. Stewart and M. 
Turner, on matters relating to the EPIC calibration and the use of the 
{\sc SAS}.  M. S. acknowledges the financial support from the Japan Society 
for the Promotion of Science (JSPS) for Young Scientists.

%
%

\bsp

\label{lastpage}


\begin{thebibliography}{99}

\bibitem[\protect\citename{Aharonian {\etal}} 2000]{Aharonian2000}
Aharonian F.A. {\etal}, 2000, ApJ, 539, 317

\bibitem[\protect\citename{Anders \& Grevesse } 1989]{Anders_G1989}
Anders E. \& Grevesse N., 1989, Geochimica et Cosmochimica Acta, 53, 197

\bibitem[\protect\citename{Baganoff {\etal}} 2001]{Baganoff2001}
Baganoff F.K. et al., 2001, Nature, 413, 45

\bibitem[\protect\citename{Bamba {\etal}} 2002]{Bamba2002}
Bamba A., Murakami H., Senda A., Takagi S., Yokogawa J., Koyama K., 2002, Proc. New Visions of the X-ray Universe in the XMM-Newton and Chandra era., in press (astro-ph/0202010)

\bibitem[\protect\citename{Coil \& Ho} 2000]{Coil2000}
Coil A.L., Ho P.T.P., 2000, ApJ, 533, 245

\bibitem[\protect\citename{Eckart {\etal}} 2002]{Eckart2002}
Eckart A., Genzel R., Ott T., Sch\"{o}del R., 2002, MNRAS, 331, 917

\bibitem[\protect\citename{Genzel {\etal}} 2000]{Genzel2000}
Genzel R., Pichon C., Eckart A., Gerhard O.E., Ott T., 2000, MNRAS, 317, 348

\bibitem[\protect\citename{Ghez {\etal}} 2000]{Ghez2000}
Ghez A.M., Morris M., Becklin E.E.,Tanner A., Kremenek T., 2000, Nature, 407, 349

\bibitem[\protect\citename{Goldwurm {\etal}} 2002]{Goldwurm2002}
Goldwurm A. {\etal}, ApJ, submitted (astro-ph/0207620)

\bibitem[\protect\citename{Ho {\etal}}1985]{Ho1985}
Ho P.T.P., Jackson J.M., Barrett A.H., Armstrong J.T., 1985, ApJ, 288, 575

\bibitem[\protect\citename{Inoue }1985]{Inoue1985}
Inoue H., 1985, Space Science Reviews, 40, 317

\bibitem[\protect\citename{Koyama }2001]{Koyama2001}
Koyama K., 2001, in Inoue H., Kunieda H., ed., ASP Conf. Ser. Vol. 251, New Century of X-ray Astronomy. Astron. Soc. Pac., San Francisco, p.50

\bibitem[\protect\citename{Koyama {\etal}}1995]{Koyama1995}
Koyama K., Petre R., Gotthelf E.V., Hwang U., Matsura M., Ozaki M., Holt, S.S., 1995, Nature, 378, 255

\bibitem[\protect\citename{Lang {\etal}}1999]{Lang1999}
Lang C.C., Morris M., Echevarria L., 1999, ApJ, 526, 727

\bibitem[\protect\citename{Lazio \& Cordes }1998]{Lazio_C1998}
Lazio T.J.W., Cordes J.M., 1998, ApJS, 118, 201

\bibitem[\protect\citename{Longair }1994]{Longair1994}
Longair M.S., 1994, High Energy Astrophysics. volume 2. second edition, Cambridge University Press, Cambridge, UK

\bibitem[\protect\citename{Mezger {\etal}} 1986]{Mezger1986}
Mezger P.G., Chini R., Kreysa E., Gem\"{u}nd H.-P., 1986, A\&A, 160, 324


\bibitem[\protect\citename{Mezger {\etal}} 1996]{Mezger1996}
Mezger P., Duschl W. J., Zylka R., 1996, A\&AR, 7, 289

\bibitem[\protect\citename{Mitsuda {\etal}} 1990]{Mitsuda1990}
Mitsuda K., Takeshima T., Kii T., Kawai N., 1990, ApJ, 353, 480

\bibitem[\protect\citename{Morris }1994]{Morris1994}
Morris M., 1994, in Genzel R., Harris A.I., ed., The nuclei of Normal Galaxies. Kluwer, Dordrecht, p.185

\bibitem[\protect\citename{Morrison \& McCammon }1983]{Morrison1983}
Morrison R., McCammon D., 1983, ApJ, 270, 119 

\bibitem[\protect\citename{LaRosa {\etal}} 2001]{LaRosa2001}
LaRosa T.N., Lazio T.J.W., Kassim N.E., 2001, ApJ, 563, 163


\bibitem[\protect\citename{Reynolds \& Keohane } 1999]{Reynolds_K1999}
Reynolds S.P., Keohane, J.W., 1999, ApJ, 525, 368
            
\bibitem[\protect\citename{Sakano {\etal}} 2002]{Sakano2002}
Sakano M., Koyama K., Murakami H., Maeda Y., Yamauchi S., 2002, ApJS, 138, 19

\bibitem[\protect\citename{Str\"{u}der {\etal}}2001]{Struder2001}
Str\"{u}der L. et al., 2001, A\&A, 365, L18

\bibitem[\protect\citename{Turner {\etal}} 2001]{Turner2001}
Turner, M.J.L. et al., 2001, A\&A, 365, L27

\bibitem[\protect\citename{Wang } 2002]{Wang2002}
Wang Q.D. 2002,  Proc. New Visions of the X-ray Universe in the XMM-Newton and Chandra era., in press (astro-ph/0202317)

\bibitem[\protect\citename{Wang {\etal}} 2002]{Wang2002}
Wang Q.D., Gotthelf E.V., Lang C.C., 2002, Nature, 415, 148

\bibitem[\protect\citename{Warwick } 2002]{Warwick2002}
Warwick R.S., 2002, Proc. New Visions of the X-ray Universe in the XMM-Newton and Chandra era., in press (astro-ph/0203333)

\bibitem[\protect\citename{Yusef-Zadeh {\etal}} 1997]{Yusef-Zadeh1997}
Yusef-Zadeh F., Wardle M., Parastaran P., 1997, ApJ, 475, 119

\bibitem[\protect\citename{Yusef-Zadeh {\etal}} 2002]{Yusef-Zadeh2002}
Yusef-Zadeh F., Law C., Wardle M., Wang Q.D., Fruscione A., Lang C.C., Cotera A., 2002, ApJ, 570, 665

\bibitem[\protect\citename{Zylka {\etal}} 1990]{Zylka1990}
Zylka R., Mezger P.G., Wink J.E., 1990, A\&A, 234, 133


\end{thebibliography}
\end{document}